\documentclass[aps,preprint,showpacs,preprintnumbers,amsmath,amssymb]{revtex4}
\usepackage{amsmath,mathrsfs,amsbsy,color,graphicx,bm,amsthm,amsfonts}
\usepackage{units}
\usepackage{bbm}
\usepackage{times}
\usepackage{dcolumn}
\usepackage{mathrsfs}
\usepackage{amsmath,amssymb,epsfig}
\usepackage{amsmath}
\newcommand{\udots}{\mathinner{\mskip1mu\raise1pt\vbox{\kern7pt\hbox{.}}
\mskip2mu\raise4pt\hbox{.}\mskip2mu\raise7pt\hbox{.}\mskip1mu}}
\begin{document}
\title{Gaussian tripartite steering in Schwarzschild black hole}
\author{Shu-Min Wu$^1$\footnote{smwu@lnnu.edu.cn}, Hao-Yu Wu$^1$, Yu-Xuan Wang$^1$, Jieci Wang$^2$\footnote{jcwang@hunnu.edu.cn (corresponding author)} }
\affiliation{$^1$ Department of Physics, Liaoning Normal University, Dalian 116029, China\\
$^2$ Department of Physics, Hunan Normal University, Changsha 410081, China}


\begin{abstract}
Multipartite steering is a fundamental quantum resource that is uniquely suited to tackling complex relativistic quantum information challenges, but its properties in the
gravitational field context remain
to be elucidated. We study the distribution of Gaussian tripartite steering in the background of  a Schwarzschild black hole. Our results show that physically accessible $2\rightarrow 1$ steerability  remains robust at any Hawking temperature, which contrasts with the behavior of $1\rightarrow 1$ steering that experiences ``sudden death" as the Hawking temperature increases, making $2\rightarrow 1$ steerability a more reliable candidate for relativistic quantum information tasks. Notably, we observe that the sudden death of quantum steering is accompanied by a peak in steering asymmetry, signifying a critical transition from two-way steering to one-way steering in the relativistic three-mode system. We find that the Hawking effect can generate inaccessible two-way steering of a bipartite system in both directions, as shown in [Phys. Rev. D {\bf93}, 125011 (2016)], while it can only generate physically inaccessible one-way $2\rightarrow 1$ steering in tripartite systems. These insights enhance our understanding of the redistribution of multipartite steering near the event horizon of the black hole.
\end{abstract}

\vspace*{0.5cm}
\maketitle
\section{introduction}
Einstein-Podolsky-Rosen (EPR) steering  \cite{L1,L2,L3}, initially  introduced by  Schr\"{o}dinger \cite{L4}, refers to the ability of one party, Alice, to remotely influence or ``steer" the state of another party, Bob, even in scenarios where Bob does not trust Alice. EPR steering can be understood as a form of quantum correlation, bridging quantum entanglement and Bell nonlocality. While it relies on quantum entanglement as a necessary resource to steer distant states, it does not always lead to the violation of Bell inequality. Unlike the symmetric nature of quantum entanglement and Bell nonlocality, EPR steering inherently displays an asymmetry: Alice's ability to steer Bob's state may differ from Bob's ability to steer Alice's. In some cases, this asymmetry can result in one-way steering, where Alice can control Bob's state, but the reverse is not possible \cite{L5}. Recently, asymmetric  steering has been successfully demonstrated in both continuous and discrete variable systems \cite{L11,L12,L13,L14,LL14}. Based on this unique feature, multipartite steering plays a significant role in quantum communication networks, and one-sided or two-sided device-independent entanglement verification, and quantum secret sharing \cite{L6,L7,L8,L9,L10,LL10}.  The advantages of multipartite steering over bipartite steering are significant, particularly in terms of the complexity, robustness, and versatility it offers \cite{L15}.

Within the framework of Einstein's theory of general relativity, the gravitational collapse of sufficiently massive stars leads to the formation of black holes in our universe. The discovery of the black hole image and gravitational waves has profoundly reshaped our understanding of black hole formation, evolution, and the physics of strong gravitational fields \cite{L16,L17,L18,L19,L20,L21,L22}. Its impact extends beyond astrophysics, reaching into the fundamental exploration of the very nature of spacetime.
The regions inside and outside the event horizon of a black hole are fundamentally distinct, giving rise to a range of unusual and mysterious phenomena. Among these is Hawking radiation, a striking consequence of quantum fluctuations near the event horizon, where particle-antiparticle pairs spontaneously form and annihilate \cite{L23,L24,L25}. This process serves as a crucial link between quantum mechanics and gravity, playing a central role in the infamous information paradox of black holes \cite{L26,L27,L28}. One promising avenue for addressing this paradox lies in the field of relativistic quantum information \cite{BRQ1,BRQ2,BRQ3,BRQ4,BRQ5,BRQ6,BRQ7,BRQ8,BRQ9,BRQ10,BRQ11,BRQ12,BRQ13,BRQ14,BRQ15,BRQ16,BRQ17,BRQ18,BRQ19,BRQ20,BRQ21,BRQ22,BRQ23,BRQ24,QQQ1,QET1,QQBJ2,QQBJ3}, which offers fresh perspectives on  quantum nonlocality, entanglement,  and coherence in a relativistic context.  Multipartite nonlocality, entanglement, and coherence have been extensively studied in a relativistic setting \cite{TRQ1,TRQ2,TRQ3,TRQ4,TRQ5,TRQ6,TRQ7,TRQ8,TRQ9,TRQ10,TRQ11}. However, the physically accessible phenomenon of multipartite steering, an even richer form of quantum correlation characterized by its inherent asymmetry, remains relatively unexplored in the gravitational background. This is one of the key reasons for studying multipartite steering in the black hole. Bipartite steering  has already been investigated in curved spacetime \cite{ST1,ST2,ST3,ST4}, with findings such as the ``sudden death"  of Gaussian steering due to the Hawking effect near the Schwarzschild black hole \cite{ST4}. Another key motivation is to explore whether Gaussian tripartite steering exhibits greater resistance to the Hawking effect compared to bipartite steering in the  Schwarzschild black hole.

In this work, we present a quantitative study on the distribution of Gaussian tripartite steering in the Schwarzschild black hole. Our study focuses on the quantum steerability shared among four key subsystems:  modes $A$ and $B$ observed by Kruskal observers Alice and Bob, respectively;  mode $C$ observed by a Schwarzschild observer Charlie, hovering outside the event horizon;  mode $\bar C$ observed by anti-Charlie, segregated by the event horizon. To achieve this, we will derive the phase-space representation for the evolution of a Gaussian tripartite state, employing the Bogoliubov transformation that relates the Kruskal vacuum to the Schwarzschild vacuum.
This framework enables us to examine two fundamental aspects. Firstly, we investigate how the Hawking effect alters quantum steerability in the directions $C\rightarrow (AB)$, $(AB)\rightarrow C$, $A\rightarrow (BC)$, and $(BC)\rightarrow A$.
Note that $C \rightarrow (AB)$ steering refers to the scenario where Charlie steers the joint system of Alice and Bob, which can be concisely denoted as $1 \rightarrow 2$ steering. Conversely, $(AB) \rightarrow C$ steering describes the case where the collective system of Alice and Bob steers Charlie, denoted as $2 \rightarrow 1$ steering.
Additionally, we assess whether Gaussian tripartite steering demonstrates greater robustness against the Hawking effect compared to bipartite $1\rightarrow 1$ steering \cite{ST4}, providing insights into the resilience of multipartite steering in extreme gravitational environment. Secondly,  we explore whether the Hawking effect can induce the generation of physically inaccessible tripartite steering, similar to accessible steering, in two directions. This sheds light on how gravitational effects influence nonlocal multipartite steering beyond direct observational reach. Our findings will reveal that Gaussian tripartite steering exhibits a richer structure and provides greater resistance to the gravitational effects of black holes compared to bipartite steering \cite{ST4}.  These properties position multipartite steering as a powerful resource for managing complex quantum information processes in curved spacetime, providing deeper insights into the behavior of multipartite quantum information near the event horizon of the black hole.

The paper is organized as follows. In Sec. II,  we provide a brief introduction to the definition and measure of Gaussian steering. In Sec. III, we describe how the Hawking effect of the black hole can be modeled as a Gaussian channel. In Sec. IV, we study the distribution of Gaussian tripartite steering in the presence of the Schwarzschild black hole. The last section is devoted to a brief conclusion.

\section{ Measure of Gaussian quantum steering  \label{GSCDGE}}
In this section, we present a concise overview of the definition and measure of Gaussian bipartite steering. For any $(n_{A} + m_{B})$-mode Gaussian state, the amplitude (position) and phase (momentum) quadratures of each mode are represented as a column vector $\hat{\xi}:=(\hat{x}^{A}_{1}, \hat{p}^{A}_{1},..., \hat{x}^{A}_{n_A}, \hat{p}^{A}_{n_A}, \hat{x}^{B}_{1}, \hat{p}^{B}_{1},..., \hat{x}^{B}_{m_B}, \hat{p}^{B}_{m_B})^{T}$, satisfying the canonical commutation relations $[\hat{x}_{j}, \hat{p}_{k}]=2i\delta_{jk}$. The properties of the state are fully characterized by its covariance matrix  ${\sigma_{AB}}$ with elements defined as $(\sigma_{AB})_{ij}=\langle\hat{\xi}_{i}\hat{\xi}_{j}+\hat{\xi}_{j}\hat{\xi}_{i}\rangle/2-\langle\hat{\xi}_{i}\rangle\langle\hat{\xi}_{j}\rangle$. The covariance matrix can be expressed in block form as $\sigma_{AB}=\left(\!\!\begin{array}{cccccccc}
A\qquad&C&\\
C^{T}\qquad&B&\\
\end{array}\!\!\right)$. Here, the sub-matrices $A$ and $B$  denote the covariance matrices  associated with the reduced states of each subsystem, respectively, and $C$
is the correlation matrix between the two parties.

The steerability from Alice to Bob via Gaussian measurements
can be quantified by \cite{L85}
\begin{eqnarray}\label{w3}
\mathcal{G}^{A\rightarrow B}(\sigma_{AB}):=\max\left\{0,-\Sigma_{j:\bar{\nu}_{j}^{AB\backslash A}<1}\ln(\bar{\nu}_{j}^{AB\backslash A})\right\},
\end{eqnarray}
where $\bar{\nu}_{j}^{AB\backslash A}(j=1,..., m_{B})$ are the symplectic eigenvalues
of the Schur complement of $A$ defined as $\bar{\sigma}_{AB\backslash A}=B-C^{T}A^{-1}C$. The quantity $\mathcal{G}^{A\rightarrow B}$  is monotonic under Gaussian local operations and classical communication  \cite{L86}, and it vanishes when Alice cannot steer Bob through Gaussian measurements. This measure has been experimentally verified using Gaussian cluster states via covariance matrix reconstruction \cite{L87}. For a specific covariance matrix $\sigma_{AB}$, the $A$ $\rightarrow$ $B$ steerability  simplifies to
\begin{eqnarray}\label{wA1}
\mathcal{G}^{A\rightarrow B}(\sigma_{AB})&=\max\left\{0,\frac{1}{2}\ln\frac{\det A}{\det \sigma_{AB}}\right\} =&\max\{0,S(A)-S(\sigma_{AB})\},
\end{eqnarray}
with $S{(\sigma)}=\frac{1}{2}\ln(\det \sigma)$ being the R\'{e}nyi-2 entropy \cite{L88}.
Furthermore, the Gaussian $B\rightarrow A$ steerability  obtained by interchanging $A$ and $B$  in the above expression is given by
\begin{eqnarray}\label{wA2}
\mathcal{G}^{B\rightarrow A}(\sigma_{AB})&=\max\left\{0,\frac{1}{2}\ln\frac{\det B}{\det \sigma_{AB}}\right\} =&\max\{0,S(B)-S(\sigma_{AB})\}.
\end{eqnarray}

\section{The quantization of scalar field  in Schwarzschild spacetime } \label{GSCDGE}
The metric of a Schwarzschild black hole \cite{Z1} can be expressed as
\begin{eqnarray}\label{l1}
ds^{2}=-(1-\frac{2M}{r})dt^{2}+(1-\frac{2M}{r})^{-1}dr^{2}+r^{2}(d\theta^{2}+\sin^{2}\theta d\varphi^{2}),
\end{eqnarray}
where $M$ denotes the mass of the Schwarzschild black hole. The dynamics of a massless scalar field are governed by the Klein-Gordon equation \cite{Z11}
\begin{eqnarray}\label{S2}
\frac{1}{\sqrt{-g}}\frac{\partial}{\partial x^{\mu}}(\sqrt{-g}g^{\mu\nu}\frac{\partial\Psi}{\partial x^{\nu}})=0.
\end{eqnarray}
The normal mode solution can be expressed as
\begin{eqnarray}\label{S3}
\Psi_{\omega lm}=\frac{1}{R(r)}\chi_{\omega l}(r)Y_{lm}(\theta,\varphi)e^{-i\omega t},
\end{eqnarray}
and the corresponding radial equation is
\begin{eqnarray}\label{S4}
\frac{d^{2}\chi_{\omega l}}{dr^{2}_{\ast}}+[\omega^{2}-V(r)]\chi_{\omega l}=0,
\end{eqnarray}
with the potential $V(r)$ defined as
\begin{eqnarray}\label{z5}
V(r)=\frac{\sqrt{f(r)h(r)}}{R(r)}\frac{d}{dr}\left[\sqrt{f(r)h(r)}\frac{dR(r)}{dr}\right]+\frac{l(l+1)f(r)}{R^{2}(r)},
\end{eqnarray}
where $Y_{lm}(\theta,\varphi)$ is a scalar spherical harmonic on the unit two-sphere. Solving Eq.(\ref{S4}) yields the incoming wave function, which is analytic across the entire spacetime
\begin{eqnarray}\label{S5}
\Psi_{in,\omega lm}=e^{-i\omega \nu}Y_{lm}(\theta,\varphi),
\end{eqnarray}
and the outgoing wave functions for regions outside and inside the event horizon are given by
\begin{eqnarray}\label{q6}
\Psi_{out,\omega lm}(r>r_{+})=e^{-i\omega \mu}Y_{lm}(\theta,\varphi),
\end{eqnarray}
\begin{eqnarray}\label{q7}
\Psi_{out,\omega lm}(r<r_{+})=e^{i\omega \mu}Y_{lm}(\theta,\varphi),
\end{eqnarray}
where $\nu=t+r_{\ast}$ and $\mu=t-r_{\ast}$. Eqs.(\ref{q6}) and (\ref{q7}) are analytic both outside and inside the event horizon, respectively, thus forming a completely orthogonal family. For second quantization in the black hole exterior, the scalar field can be expanded as
\begin{eqnarray}\label{q8}
\Phi_{out}&=&\sum_{lm}\int d\omega[b_{in,\omega lm}\Psi_{out,\omega lm}(r<r_{+})
+b_{in,\omega lm}^{\dag}\Psi_{out,\omega lm}^{\ast}(r<r_{+})\notag\\
&+&b_{out,\omega lm}\Psi_{out,\omega lm}(r>r_{+})+b_{out,\omega lm}^{\dag}\Psi_{out,\omega lm}^{\ast}(r>r_{+})],
\end{eqnarray}
where $b_{in,\omega lm}$ and $b_{in,\omega lm}^{\dagger}$ are the annihilation and creation operators acting on the vacuum of the interior region of the black hole, while $b_{out,\omega lm}$ and $b_{out,\omega lm}^{\dagger}$ correspond to the exterior region \cite{BRQ11}. Thus, the Schwarzschild Fock vacuum state can be defined as
\begin{eqnarray}\label{S9}
b_{in,\omega lm}|0\rangle_{in}=b_{out,\omega lm}|0\rangle_{out}=0.
\end{eqnarray}

The lightlike Kruskal coordinates are defined as
\begin{eqnarray}\label{Z1}
U&=&4Me^{-\frac{\mu}{4M}},V=4Me^{\frac{\nu}{4M}},\quad \mathrm {if}\ r<r_{+};\notag\\
U&=&-4Me^{-\frac{\mu}{4M}},V=4Me^{\frac{\nu}{4M}},\ \mathrm {if}\ r>r_{+}.
\end{eqnarray}
Using these coordinates, the Schwarzschild modes can be rewritten as
\begin{eqnarray}\label{Z2}
\Psi_{out,\omega lm}(r<r_{+})=e^{-4i\omega M\ln[-\frac{U}{4M}]}Y_{lm}(\theta,\varphi),
\end{eqnarray}
\begin{eqnarray}\label{Z3}
\Psi_{out,\omega lm}(r>r_{+})=e^{4i\omega M\ln[\frac{U}{4M}]}Y_{lm}(\theta,\varphi).
\end{eqnarray}
Following Ruffini and Damour \cite{Z3}, the outgoing modes in Kruskal spacetime are expressed as
\begin{eqnarray}\label{S10}
\Psi_{I,\omega lm}=e^{(\pi\omega/2\kappa)}\Psi_{out,\omega lm}(r>r_{+})
+e^{-(\pi\omega/2\kappa)}\Psi_{out,\omega lm}^{\ast}(r<r_{+}),
\end{eqnarray}
\begin{eqnarray}\label{S11}
\Psi_{II,\omega lm}=e^{(-\pi\omega/2\kappa)}\Psi_{out,\omega lm}^{\ast}(r>r_{+})
+e^{(\pi\omega/2\kappa)}\Psi_{out,\omega lm}(r<r_{+}).
\end{eqnarray}
Thus,  the scalar field $\Phi_{out}$ in terms of $\Psi_{I,\omega lm}$ and $\Psi_{II,\omega lm}$ in  Kruskal spacetime is quantized as
\begin{eqnarray}\label{S12}
\Phi_{out}&=&\sum_{lm} \int d\omega[2\sinh{(\pi\omega/\kappa)]^{-1/2}}[a_{out,\omega lm}\Psi_{I,\omega lm}\notag\\
&+&a_{out,\omega lm}^{\dagger}\Psi_{I,\omega lm}^{\ast}+a_{in,\omega lm}\Psi_{II,\omega lm}+a_{in,\omega lm}^{\dagger}\Psi_{II,\omega lm}^{\ast}].
\end{eqnarray}
Here, the annihilation operator $a_{out,\omega lm}$ defines the Kruskal vacuum
\begin{eqnarray}\label{S13}
a_{out,\omega lm}|0\rangle_{K}=0.
\end{eqnarray}
From Eqs.(\ref{q8}) and (\ref{S12}), the Bogoliubov relations for the annihilation and creation operators are
\begin{eqnarray}\label{S14}
a_{out,\omega lm}=\frac{b_{out,\omega lm}}{\sqrt{1-e^{-\omega/T}}}-\frac{b_{in,\omega lm}^{\dag}}{\sqrt{e^{\omega/T}-1}},
\end{eqnarray}
\begin{eqnarray}\label{S15}
a_{out,\omega lm}^{\dag}=\frac{b_{out,\omega lm}^{\dag}}{\sqrt{1-e^{-\omega/T}}}
-\frac{b_{in,\omega lm}}{\sqrt{e^{\omega/T}-1}},
\end{eqnarray}
where $T=\frac{1}{8\pi M}$ represents Hawking temperature \cite{BRQ11}.

Upon properly normalizing the state vector, the Kruskal vacuum in the Schwarzschild black hole can be expressed as a maximally entangled two-mode squeezed state
 \begin{eqnarray}\label{S23}
|0_{\omega}\rangle_{K}=\frac{1}{\cosh \eta}\underset{n=0}{\overset{\infty}{\sum}}\tanh^{n}\eta|n_{\omega}\rangle_{out}|n_{\omega}\rangle_{in}=\hat{U}|0\rangle_{out}|0\rangle_{in},
\end{eqnarray}
where $\cosh \eta=\frac{1}{\sqrt{1-e^{-\omega/T}}}$, and  $\hat{U}=\exp[\eta(b^{\dag}_{out}b^{\dag}_{in}-b_{out}b_{in})]$ represents the two-mode squeezing operator. Here, the $\eta$ is the Hawking temperature parameter.
This operator $\hat{U}$ is a Gaussian transformation, preserving the Gaussian nature of the states.  The symplectic phase-space representation of the two-mode squeezing operation $\hat{U}$ has the form
\begin{eqnarray}\label{S25}
S_{C,\bar{C}}(\eta)= \left(
\begin{array}{cccc}
\cosh \eta&0& \sinh \eta&0\\
0&\cosh \eta&0&-\sinh \eta\\
\sinh \eta&0&\cosh \eta&0\\
0&-\sinh \eta&0&\cosh \eta
 \end{array}
 \right).
\end{eqnarray}

\section{Distribution of Gaussian tripartite steering in the presence of the Schwarzschild black hole \label{GSCDGE}}
The three-mode pure Gaussian state shared by Alice, Bob, and Charlie in an asymptotically flat spacetime is characterized by the covariance matrix  \cite{Z6,Z21}
\begin{eqnarray}\label{S26}
\sigma_{ABC}(s)=\left(\begin{array}{ccc}
\sigma_{A}(s)&\zeta(s)&\zeta(s)\\
\zeta(s)&\sigma_{B}(s)&\zeta(s)\\
\zeta(s)&\zeta(s)&\sigma_{C}(s)\\
\end{array}\right),
\end{eqnarray}
where the sub-matrices are defined as
\begin{eqnarray}\label{S6}
\sigma_{A}(s)=\sigma_{B}(s)=\sigma_{C}(s)=\left(\begin{array}{ccc}
b\qquad 0\\
0\qquad b\\
\end{array}\right),
\qquad\zeta(s)=\left(\begin{array}{ccc}
z_{1}\qquad 0\\
0\qquad z_{2}\\
\end{array}\right).
\end{eqnarray}
The
parameters $b$, $z_{1}$, and $z_{2}$ depend on the squeezing parameter $s$ as
\begin{eqnarray}\label{S7}
b=\frac{1}{3}\sqrt{4\cosh(4s)+5},
\end{eqnarray}
\begin{eqnarray}\label{S8}
z_{1}=\frac{2\sinh^{2}(2s)+3\sinh(4s)}{3\sqrt{4\cosh(4s)+5}},
\end{eqnarray}
\begin{eqnarray}\label{S9}
z_{2}=\frac{2\sinh^{2}(2s)-3\sinh(4s)}{3\sqrt{4\cosh(4s)+5}}.
\end{eqnarray}
In this setup, Charlie moves toward the event horizon of the black hole, while Alice and Bob remain stationary in an asymptotically flat region.
As Charlie approaches the event horizon, the transition from the Kruskal mode to the Schwarzschild mode introduces a symplectic transformation denoted as $S_{C\bar{C}}$, in the phase space. This transformation maps Charlie's mode $C$ into two components: mode
 $C$, observed outside the black hole, and mode
$\bar C$, situated inside the event horizon. Thus, a comprehensive description of the Gaussian system consists of four modes:  modes $A$ and $B$ are observed by Kruskal observers Alice and Bob, respectively; mode $C$ is observed by a Schwarzschild observer, Charlie;  mode $\bar C$ is observed by anti-Charlie, who is separated by the event horizon. The covariance matrix characterizing this four-mode Gaussian state is expressed as \cite{BRQ7}
\begin{eqnarray}\label{S27}
\nonumber\sigma_{ABC\bar C}(s,\eta) &=& \big[I_{AB}\oplus  S_{C, \bar C}(\eta)\big] \big[\sigma_{ABC}(s)\oplus I_{\bar C}\big]\\&& \big[I_{AB}\oplus  S_{C, \bar C}(\eta)\big]^T\,\nonumber\\
 &=& \left(
\begin{array}{cccc}
\mathcal{\sigma}_{A}(s) & \mathcal{E}_{AB}(s) & \mathcal{E}_{AC}(s,\eta) & \mathcal{E}_{A\bar C}(s,\eta) \\
\mathcal{E}^{\sf T}_{AB}(s) &  \mathcal{\sigma}_{B}(s) & \mathcal{E}_{BC}(s,\eta) & \mathcal{E}_{B\bar C}(s,\eta) \\
\mathcal{E}^{\sf T}_{AC}(s,\eta) & \mathcal{E}^{\sf T}_{BC}(s,\eta) & \mathcal{\sigma}_{C}(s,\eta) & \mathcal{E}_{C\bar C}(s,\eta)\\
\mathcal{E}^{\sf T}_{A\bar C}(s,\eta) & \mathcal{E}^{\sf T}_{B\bar C}(s,\eta) & \mathcal{E}^{\sf T}_{C\bar C}(s,\eta) &  \mathcal{\sigma}_{\bar C}(s,\eta)
\end{array}
\right)\,,
\end{eqnarray}
where $\big[\sigma_{ABC}(s)\oplus I_{\bar C}\big]$ is the initial covariance matrix for the entire Gaussian state.
The diagonal elements in Eq.(\ref{S27}) take the forms
\begin{gather}
\mathcal{\sigma}_{C}(s,\eta)=[b\cosh^{2}\eta + \sinh^{2}\eta]I_2,\\ \nonumber\mathcal{\sigma}_{\bar C}(s,\eta)=[\cosh^{2}\eta + b\sinh^{2}\eta]I_2,\nonumber
\end{gather}
where $I_2$ denotes the unity matrix in $2\times2$ space.
The non-diagonal elements take the following forms
\begin{gather}\nonumber\mathcal{E}_{AB}(s)=\left(\begin{array}{ccc}
z_{1}\qquad 0\\
0\qquad z_{2}\\
\end{array}\right),\end{gather}
\begin{gather}\mathcal{E}_{BC}(s,\eta)=\mathcal{E}_{AC}(s,\eta)=\left(\begin{array}{ccc}
z_{1}\cosh \eta\qquad 0\\
0\qquad z_{2}\cosh \eta\\
\end{array}\right),\end{gather}
\begin{gather}\nonumber\mathcal{E}_{A\bar C}(s,\eta)=\mathcal{E}_{B\bar C}(s,\eta)=\left(\begin{array}{ccc}
z_{1}\sinh \eta\qquad 0\\
0\qquad -z_{2}\sinh \eta\\
\end{array}\right),\end{gather} and \begin{gather}\nonumber\mathcal{E}_{C\bar C}(s,\eta)=\left(\begin{array}{ccc}
(1+b)\cosh\eta\sinh\eta\qquad 0\\
0\qquad (1+b)\cosh\eta\sinh\eta\\
\end{array}\right).\nonumber \end{gather}

\subsection{Gaussian tripartite steering between initially correlated modes}
Because the exterior region of the black hole is causally disconnected from the inner region, Alice, Bob, and Charlie cannot approach mode $\bar{C}$. Consequently, the covariance matrix for the three observable modes $A$, $B$, and $C$  is obtained by tracing out mode $\bar{C}$, and is expressed as
\begin{eqnarray}\label{B2}
\sigma_{ABC}(s,\eta) &=&
\left(\begin{array}{cccc}
\mathcal{\sigma}_{A}(s) & \mathcal{E}_{AB}(s) & \mathcal{E}_{AC}(s,\eta)  \\
\mathcal{E}^{\sf T}_{AB}(s) &  \mathcal{\sigma}_{B}(s) & \mathcal{E}_{BC}(s,\eta) &  \\
\mathcal{E}^{\sf T}_{AC}(s,\eta) & \mathcal{E}^{\sf T}_{BC}(s,\eta) & \mathcal{\sigma}_{C}(s,\eta) & \\
\end{array}
\right)\,.
\end{eqnarray}
Using this covariance matrix $\sigma_{ABC}(s,\eta)$ and Eqs.(\ref{wA1}) and (\ref{wA2}),  we derive the analytic expressions of Gaussian tripartite steering for various $2\rightarrow 1$ and  $1\rightarrow 2$ partitions of the state  given by Eq.(\ref{B2}). These steering quantifiers are given as
\begin{eqnarray}\label{S31}
\mathcal{G}^{A \rightarrow\ (BC)}(\sigma_{ABC})&=&\max\left\{0,\frac{1}{2}\ln\frac{\frac{1}{9}[{5+4\cosh(4s)}]}{{\rm det}\sigma_{ABC}}\right\},
\end{eqnarray}
\begin{eqnarray}\label{S32}
\mathcal{G}^{(BC) \rightarrow\ A}(\sigma_{ABC})&=&\max\left\{0,\frac{1}{2}\ln\frac{{\rm det}\sigma_{BC}}{{\rm det}\sigma_{ABC}}\right\},
\end{eqnarray}
\begin{eqnarray}\label{S33}
\mathcal{G}^{(AB) \rightarrow\ C}(\sigma_{ABC})&=&\max\left\{0,\frac{1}{2}\ln\frac{\frac{1}{9}[5+4\cosh(4s)]}{{\rm det} \sigma_{ABC}}\right\},
\end{eqnarray}
\begin{eqnarray}\label{S34}
\mathcal{G}^{C \rightarrow\ (AB)}(\sigma_{ABC})&=&\max\left\{0,\frac{1}{2}\ln\frac{[\frac{1}{3}\sqrt{5+4\cosh(4s)}\cosh^{2}(\eta)+\sinh^{2}(\eta)]^{2}}{{\rm det}\sigma_{ABC}}\right\},
\end{eqnarray}
where
\begin{eqnarray}\label{S313}
\nonumber \rm{det}\sigma_{ABC}&=&\frac{1}{36}\{21+8\cosh(2\eta)-4\cosh[2(\eta-2s)]+\cosh[4(\eta-s)] \\ \nonumber
&+&6\cosh(4s)
-3\sqrt{5+4\cosh(4s)}  +\cosh(4\eta)(7+3\sqrt{5+4\cosh(4s)}) \\ \nonumber
&+&\cosh[4(\eta+s)]-4\cosh[2(\eta+2s)]\}  \nonumber,
\\\nonumber \qquad \quad \rm{det}\sigma_{BC}&=&\frac{1}{9[5+4\cosh(4s)]^{2}}\{[5+4\cosh(4s)]^{\frac{3}{2}}\sinh^{2}(\eta) +\cosh^{2}(\eta)[3\cosh(2s)\\ \nonumber
&-&\sinh(2s)]^{3}[\cosh(2s)+\sinh(2s)]\}\{[5+4\cosh(4s)]^{\frac{3}{2}}\sinh^{2}(\eta)\\ \nonumber
&+&\cosh^{2}(\eta)[\cosh(2s)-\sinh(2s)][3\cosh(2s)+\sinh(2s)]^{3}\}  \nonumber.
\end{eqnarray}
From the above expressions, it is evident that Gaussian tripartite steering is influenced not only by the initial squeezing parameter $s$, which encodes the quantum resources in the system, but also by the Hawking temperature parameter
$\eta$, which encapsulates the thermal effects of the black hole. The dependency on
$\eta$ reflects the sensitivity of Gaussian tripartite steering to the thermal dynamics near the event horizon. It should be noted that since modes $A$ and $B$ are symmetrical, we obtain the  expressions as $\mathcal{G}^{B \rightarrow\ (AC)}=\mathcal{G}^{A \rightarrow\ (BC)}$
and $\mathcal{G}^{(AC) \rightarrow\ B}=\mathcal{G}^{(BC) \rightarrow\ A}$.

\begin{figure}
\begin{minipage}[t]{0.5\linewidth}
\centering
\includegraphics[width=3.0in,height=5.3cm]{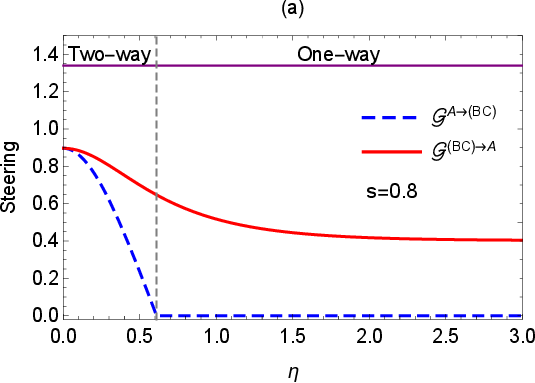}
\label{fig1a}
\end{minipage}%
\begin{minipage}[t]{0.5\linewidth}
\centering
\includegraphics[width=3.0in,height=5.3cm]{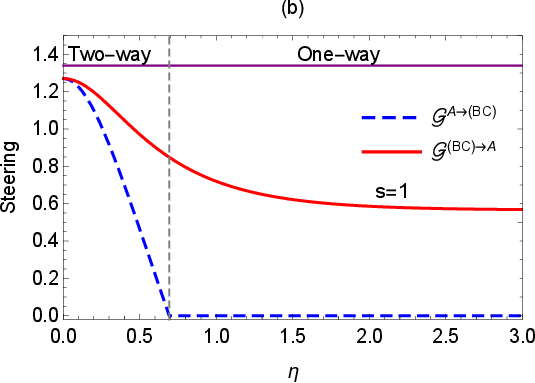}
\label{fig1c}
\end{minipage}%

\begin{minipage}[t]{0.5\linewidth}
\centering
\includegraphics[width=3.0in,height=5.3cm]{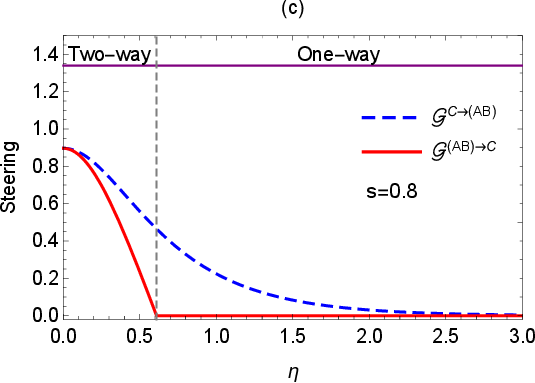}
\label{fig1a}
\end{minipage}%
\begin{minipage}[t]{0.5\linewidth}
\centering
\includegraphics[width=3.0in,height=5.3cm]{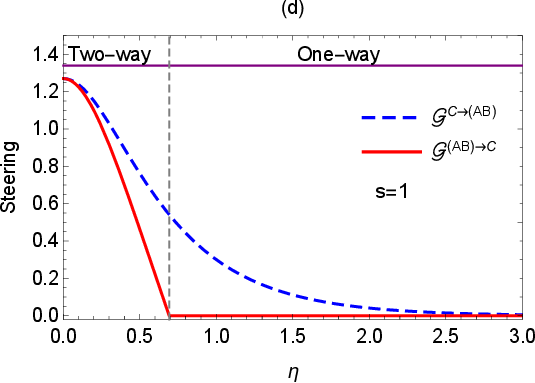}
\label{fig1c}
\end{minipage}%
\caption{$\mathcal{G}^{A \rightarrow\ (BC)}$, $\mathcal{G}^{(BC) \rightarrow\ A}$, $\mathcal{G}^{C \rightarrow\ (AB)}$, and $\mathcal{G}^{(AB) \rightarrow\ C}$ as functions of the Hawking temperature parameter $\eta$ for different initial parameters $s$.}
\label{F1}
\end{figure}

In Fig.\ref{F1},  we present a detailed analysis of Gaussian tripartite steering for various $2\rightarrow 1$ and  $1\rightarrow 2$ partitions of the initially correlated modes between Alice, Bob, and Charlie, plotted as functions of the Hawking temperature parameter $\eta$ of the Schwarzschild black hole. Note that the relation between the parameter $\eta$ and Hawking temperature $T$ is given by $\cosh \eta=\frac{1}{\sqrt{1-e^{-\omega/T}}}$, indicating that the $\eta$ is a monotonically increasing function of the $T$  for a fixed $\omega$.  From Fig.\ref{F1}, we observe that  $\mathcal{G}^{(BC) \rightarrow\ A}$ first decreases and then approaches an asymptotic value with the growth of the Hawking temperature parameter $\eta$, which contrasts sharply with the behavior of Gaussian bipartite $1 \rightarrow 1$ steering in curved spacetime, where bipartite steering suffers ``sudden death" with the $\eta$  \cite{ST4}. This suggests that Gaussian tripartite steering exhibits a greater resilience to the Hawking effect of the black hole, providing a more robust form of quantum correlation than bipartite steering. In addition, $\mathcal{G}^{A \rightarrow\ (BC)}$ experiences ``sudden death" under the influence of the Hawking effect, showing a transition in the initial system from two-way steering to one-way steering. From Fig.\ref{F1}(a) and (b), we find that the point of ``sudden death" of $\mathcal{G}^{A \rightarrow\ (BC)}$ and the asymptotic value of $\mathcal{G}^{(BC) \rightarrow\ A}$ depend on the initial squeezing parameter $s$,
highlighting the crucial role that the quantum resources shared in the initial state play in Gaussian tripartite steering. Moreover,  from Fig.\ref{F1}(c) and (d), we also find that $\mathcal{G}^{(AB) \rightarrow\ C}$ undergoes ``sudden death" with the $\eta$, while $\mathcal{G}^{C \rightarrow\ (AB)}$ vanishes only at the extreme black hole limit ($\eta\rightarrow \infty$). This suggests an intriguing asymmetry: the Schwarzschild mode steering the Kruskal modes is more easily affected by the Hawking effect than the reverse process, where the Kruskal modes attempt to steer the Schwarzschild mode.

Unlike quantum entanglement, quantum steering displays a distinctive asymmetry, a feature that has been experimentally verified in recent studies \cite{L11,L12,L13,L14,LL14}. In asymptotically flat regions, Gaussian tripartite steerability remains perfectly symmetrical. However, the Hawking effect of the black hole disrupts this symmetry. For instance, as illustrated in Fig.\ref{F1},  $\mathcal{G}^{(BC) \rightarrow\ A}$ is greater than $\mathcal{G}^{A \rightarrow\ (BC)}$, and $\mathcal{G}^{C \rightarrow\ (AB)}$ exceeds  $\mathcal{G}^{(AB) \rightarrow\ C}$.
These observations highlight a clear asymmetry in the steering dynamics within the black hole environment. To quantify the difference between  $\mathcal{G}^{(BC) \rightarrow\ A}$  and  $\mathcal{G}^{A \rightarrow\ (BC)}$,  we define the Gaussian steering asymmetry $\mathcal{G}^{\Delta}_{A|BC}$ as
\begin{eqnarray}
\mathcal{G}^{\Delta}_{A|BC}=|\mathcal{G}^{A\rightarrow (BC)}-\mathcal{G}^{(BC)\rightarrow A}|.
\end{eqnarray}
Similarly, to measure the asymmetry between $\mathcal{G}^{(AB) \rightarrow\ C}$ and $\mathcal{G}^{C\rightarrow (AB)}$, we introduce the Gaussian steering asymmetry $\mathcal{G}^{\Delta}_{AB|C}$,   defined as
\begin{eqnarray}
\mathcal{G}^{\Delta}_{AB|C}=|\mathcal{G}^{(AB)\rightarrow C}-\mathcal{G}^{C\rightarrow (AB)}|.
\end{eqnarray}
By examining these asymmetries, we gain insight into the underlying physical mechanisms that govern the behavior of Gaussian tripartite steering near the event horizon of the black hole.

\begin{figure}[htbp]
\centering
\includegraphics[height=2.5in,width=3.0in]{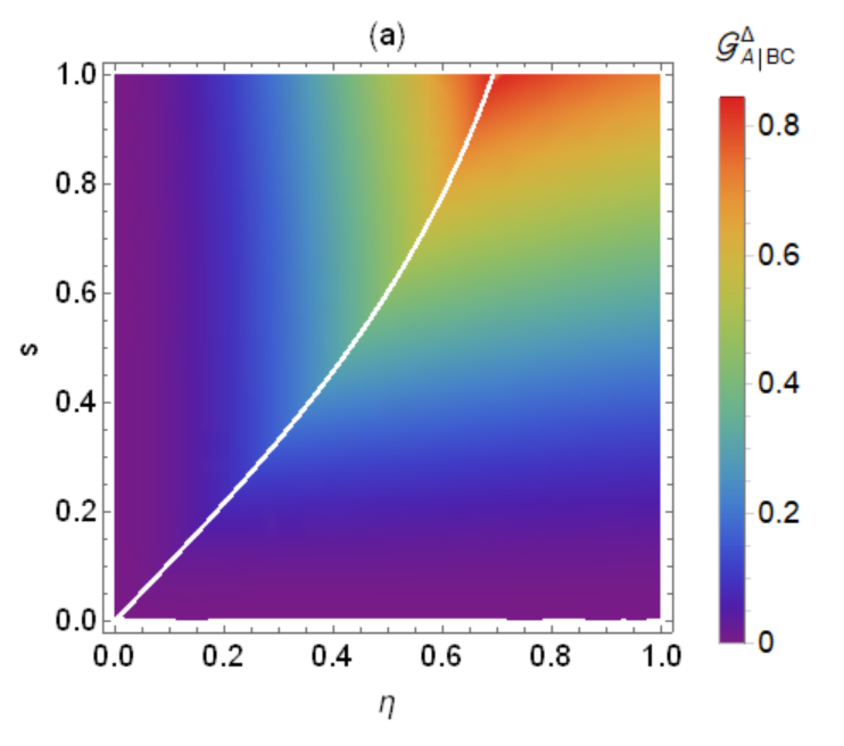}
\includegraphics[height=2.5in,width=3.0in]{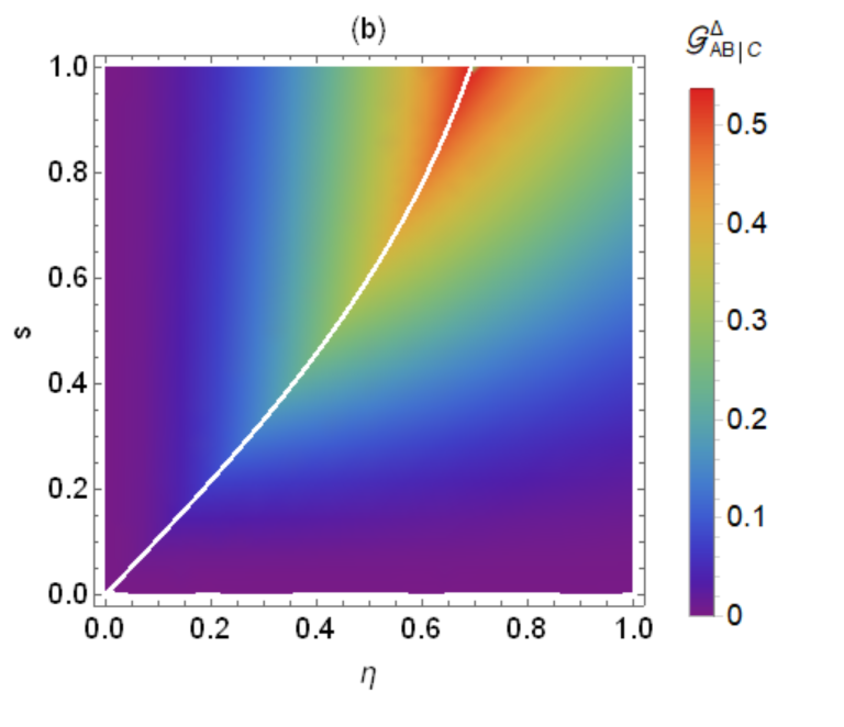}
\caption{The Gaussian tripartite steering asymmetry $\mathcal{G}^{\Delta}_{A|BC}$ and $\mathcal{G}^{\Delta}_{AB|C}$  as functions
of the Hawking temperature parameter $\eta$ and the squeezing parameter $s$.}\label{F2}
\end{figure}

In Fig.\ref{F2}, we plot the Gaussian tripartite steering asymmetry $\mathcal{G}^{\Delta}_{A|BC}$ and $\mathcal{G}^{\Delta}_{AB|C}$ as functions of the Hawking temperature parameter $\eta$ and the squeezing parameter $s$. In accordance with Eqs.(\ref{S31}) and (\ref{S32}), the steering asymmetry for one-way steering between  $\mathcal{G}^{(BC) \rightarrow\ A}$  and  $\mathcal{G}^{A \rightarrow\ (BC)}$ is determined by
\begin{eqnarray}
\nonumber\mathcal{G}^{\Delta,1}_{A|BC}=\max\left\{0,\frac{1}{2}\ln\frac{{\rm det}\sigma_{BC}}{{\rm det}\sigma_{ABC}}\right\},\nonumber
\end{eqnarray}
which is a decreasing function of the $\eta$. In contrast, the steering asymmetry for two-way steering is described by
\begin{eqnarray}
\nonumber \mathcal{G}^{\Delta,2}_{A|BC}=\max\left\{0,\frac{1}{2}\ln\frac{{\rm det}\sigma_{BC}}{\frac{1}{9}[{5+4\cosh(4s)]}}\right\},\nonumber
\end{eqnarray}
which is an increasing function of the $\eta$.
The maximal steering asymmetry occurs at the transition point from two-way steering to one-way steering in curved spacetime. Through detailed calculation, we find that the configuration of parameters that maximizes the steering asymmetry is
$$\eta_0=\rm{arccosh}\left[-\frac{\sqrt{[5+4\cosh(4s)-3\sqrt{5+4\cosh(4s)}]\sinh^{2}(2s)}}{2\sqrt{\sinh^{4}(2s)}}\right].$$ This condition is identical to the scenario in which the
$A\rightarrow (BC)$ steering undergoes ``sudden death" with the $\eta$.
In addition, from Eqs.(\ref{S33}) and (\ref{S34}),  the steering asymmetry for one-way steering between  $\mathcal{G}^{(AB) \rightarrow\ C}$  and  $\mathcal{G}^{C \rightarrow\ (AB)}$
is determined by
\begin{eqnarray}
\nonumber \mathcal{G}^{\Delta,1}_{AB|C}=\max\left\{0,\frac{1}{2}\ln\frac{[\frac{1}{3}\sqrt{5+4\cosh(4s)}\cosh^{2}(\eta)+\sinh^{2}(\eta)]^{2}}{{\rm det}\sigma_{ABC}}\right\},
\nonumber\end{eqnarray}
while the two-way steering asymmetry is
\begin{eqnarray}
\nonumber \mathcal{G}^{\Delta,2}_{AB|C}=\max\left\{0,\frac{1}{2}\ln\frac{[\frac{1}{3}\sqrt{5+4\cosh(4s)}\cosh^{2}(\eta)+\sinh^{2}(\eta)]^{2}}{\frac{1}{9}[5+4\cosh(4s)]}\right\}.\nonumber
\end{eqnarray}
Interestingly, the maximal steering asymmetry $\mathcal{G}^{\Delta}_{AB|C}$ is also point $\eta_0$, since $\mathcal{G}^{A \rightarrow\ (BC)}$ is equal to $\mathcal{G}^{(AB) \rightarrow\ C}$ in curved spacetime.
In Fig.\ref{F2},  we observe that the value of  $\eta_0$ for maximal steering asymmetry increases as the squeezing parameter $s$  grows. This suggests that the quantum resources shared in the initial state characterized by the squeezing parameter play a crucial role in the Gaussian steering asymmetry in relativistic environment. This indicates a deep connection between the quantum state preparation and the influence of the Hawking effect on Gaussian tripartite steering, emphasizing the importance of both the Hawking temperature and squeezing in controlling quantum correlations in curved spacetime.

\subsection{Generation of Gaussian tripartite steering between initially uncorrelated modes }

In this subsection, we explore the phenomenon of Gaussian tripartite steering among the initially uncorrelated modes in the Schwarzschild black hole. By tracing over mode $B$, we derive the covariance matrix between Alice, Charlie, and anti-Charlie, which is given by the following expression
\begin{eqnarray}\label{B3}
\sigma_{AC\bar C}(s,\eta) &=&
\left(\begin{array}{cccc}
\mathcal{\sigma}_{A}(s) & \mathcal{E}_{AC}(s,\eta) & \mathcal{E}_{A\bar C}(s,\eta) \\
\mathcal{E}^{\sf T}_{AC}(s,\eta) & \mathcal{\sigma}_{C} (s,\eta)& \mathcal{E}_{C\bar C}(s,\eta) \\
\mathcal{E}^{\sf T}_{A\bar C}(s,\eta) & \mathcal{E}^{\sf T}_{C\bar C}(s,\eta) &  \mathcal{\sigma}_{\bar C}(s,\eta)
\end{array}
\right)\,.
\end{eqnarray}
Employing Eqs.(\ref{wA1}),  (\ref{wA2}), and  (\ref{B3}), the analytic expressions of the Gaussian steering among the modes $A$,  $C$, and $\bar C$ read
\begin{eqnarray}\label{S39}
\mathcal{G}^{(A\bar C) \rightarrow\ C}(\sigma_{AC\bar{C}})&=&\mathcal{G}^{(AC) \rightarrow\ \bar{C}}(\sigma_{AC\bar{C}})=\max\left\{0,\frac{1}{2}\ln\frac{{\rm det}\sigma_{AC}}{\frac{1}{9}\cosh^{2}(2\eta)[5+4\cosh(4s)]}\right\},
\\\mathcal{G}^{C \rightarrow\ (A\bar C)}(\sigma_{AC\bar{C}})&=&\max\left\{0,\frac{1}{2}\ln\frac{[\frac{1}{3}\sqrt{5+4\cosh(4s)}\cosh^{2}(\eta)+\sinh^{2}(\eta)]^{2}}{\frac{1}{9}\cosh^{2}(2\eta)[5+4\cosh(4s)]}\right\}=0,
\\\mathcal{G}^{\bar{C} \rightarrow\ (AC)}(\sigma_{AC\bar{C}})&=&\max\left\{0,\frac{1}{2}\ln\frac{[\frac{1}{3}\sqrt{5+4\cosh(4s)}\sinh^{2}(\eta)+\cosh^{2}(\eta)]^{2}}{\frac{1}{9}\cosh^{2}(2\eta)[5+4\cosh(4s)]}\right\}=0,
\\\mathcal{G}^{A \rightarrow\ (C\bar{C})}(\sigma_{AC\bar{C}})&=&\max\left\{0,\frac{1}{2}\ln\frac{\frac{1}{9}[5+4\cosh(4s)]}{\frac{1}{9}\cosh^{2}(2\eta)[5+4\cosh(4s)]}\right\}=0,
\\\mathcal{G}^{(C\bar{C}) \rightarrow\ A}(\sigma_{AC\bar{C}})&=&\max\left\{0,\frac{1}{2}\ln\frac{\frac{1}{9}\cosh^{2}(2\eta)[5+4\cosh(4s)]}{\frac{1}{9}\cosh^{2}(2\eta)[5+4\cosh(4s)]}\right\}=0,
\end{eqnarray}
where $\rm{det}\sigma_{AC}=\rm{det}\sigma_{BC}$.

Next, by tracing over mode $C$, we obtain the covariance matrix between Alice, Bob, and anti-Charlie, which is expressed as
\begin{eqnarray}\label{B33}
\sigma_{AB\bar C}(s,\eta) &=&
\left(\begin{array}{cccc}
\mathcal{\sigma}_{A}(s) & \mathcal{E}_{AB}(s) & \mathcal{E}_{A\bar C}(s,\eta) \\
\mathcal{E}^{\sf T}_{AB}(s) &  \mathcal{\sigma}_{B}(s) &\mathcal{E}_{B\bar C}(s,\eta) \\
\mathcal{E}^{\sf T}_{A\bar C}(s,\eta) & \mathcal{E}^{\sf T}_{B\bar C}(s,\eta) &\mathcal{\sigma}_{\bar C}(s,\eta)
\end{array}
\right)\,.
\end{eqnarray}
\begin{figure}
\begin{minipage}[t]{0.5\linewidth}
\centering
\includegraphics[width=3.0in,height=5.3cm]{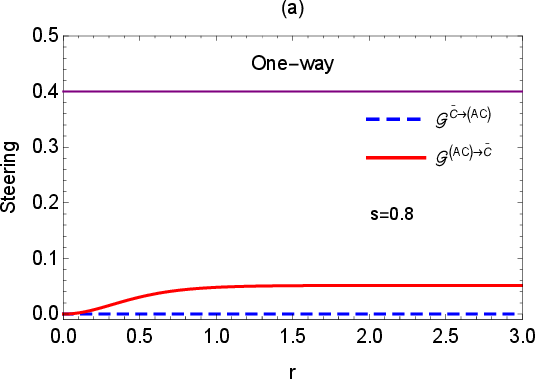}
\label{fig1a}
\end{minipage}%
\begin{minipage}[t]{0.5\linewidth}
\centering
\includegraphics[width=3.0in,height=5.3cm]{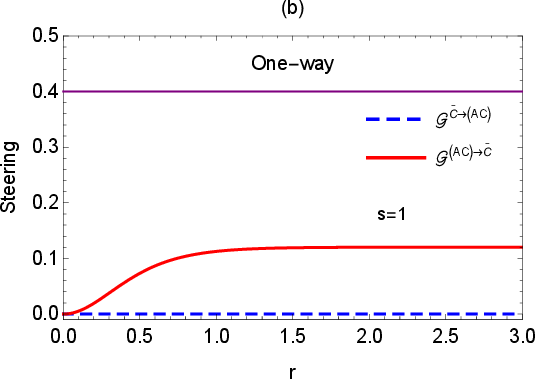}
\label{fig1c}
\end{minipage}%

\begin{minipage}[t]{0.5\linewidth}
\centering
\includegraphics[width=3.0in,height=5.3cm]{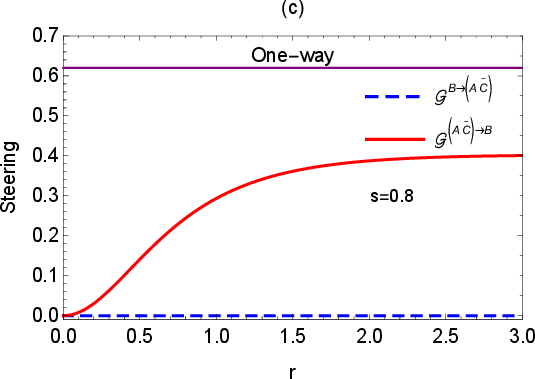}
\label{fig1a}
\end{minipage}%
\begin{minipage}[t]{0.5\linewidth}
\centering
\includegraphics[width=3.0in,height=5.3cm]{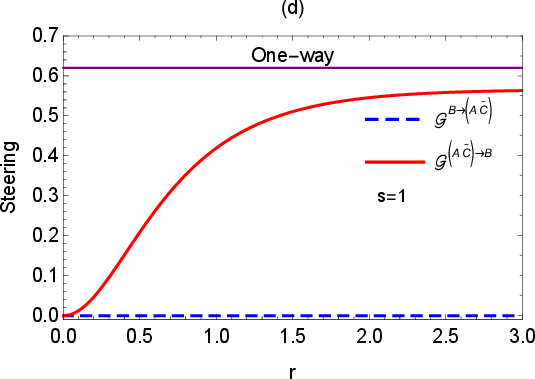}
\label{fig1c}
\end{minipage}%
\caption{$\mathcal{G}^{\bar C \rightarrow\ (AC)}$, $\mathcal{G}^{(AC) \rightarrow\ \bar C}$, $\mathcal{G}^{B \rightarrow\ (A \bar C)}$, and $\mathcal{G}^{(A\bar C) \rightarrow\ B}$ as functions of the Hawking temperature parameter $\eta$ for fixed $s=0.8$ and $s=1$, respectively. }
\label{F3}
\end{figure}
The corresponding analytic expressions of the Gaussian steering among modes $A$, $B$, and  $\bar C$ are found to be
\begin{eqnarray}\label{SS39}
\mathcal{G}^{(A\bar C) \rightarrow\ B}(\sigma_{AB\bar{C}})&=&\mathcal{G}^{(B\bar C) \rightarrow\ A}(\sigma_{AB\bar{C}})=\max\left\{0,\frac{1}{2}\ln\frac{{\rm det}\sigma_{AC}}{{\rm det}\sigma_{AB\bar{C}}}\right\},
\\\mathcal{G}^{A \rightarrow\ (B\bar C)}(\sigma_{AB\bar{C}})&=&\mathcal{G}^{B \rightarrow\ (A\bar C)}(\sigma_{AB\bar{C}})=\max\left\{0,\frac{1}{2}\ln\frac{\frac{1}{9}[{5+4\cosh(4s)}]}{{\rm det}\sigma_{AB\bar{C}}}\right\}=0,
\\\mathcal{G}^{(AB) \rightarrow\ \bar{C} }(\sigma_{AB\bar{C}})&=&\max\left\{0,\frac{1}{2}\ln\frac{\frac{1}{9}[5+4\cosh(4s)]}{{\rm det}\sigma_{AB\bar{C}}}\right\}=0,
\\\mathcal{G}^{\bar{C} \rightarrow\ (AB)}(\sigma_{AB\bar{C}})&=&\max\left\{0,\frac{1}{2}\ln\frac{[\frac{1}{3}\sqrt{5+4\cosh(4s)}\sinh^{2}(\eta)+\cosh^{2}(\eta)]^{2}}{{\rm det}\sigma_{AB\bar{C}}}\right\}=0,
\end{eqnarray}
where the determinant $\rm{det}\sigma_{AB\bar{C}}$  is given by
\begin{eqnarray}\label{S313}
\nonumber \rm{det}\sigma_{AB\bar{C}}&=&\frac{1}{36}\{21-8\cosh(2\eta)+4\cosh[2(\eta-2s)]+\cosh[4(\eta-s)] \\ \nonumber
&+&6\cosh(4s)
-3\sqrt{5+4\cosh(4s)} +\cosh(4\eta)(7+3\sqrt{5+4\cosh(4s)}) \\ \nonumber
&+&\cosh[4(\eta+s)]+4\cosh[2(\eta+2s)]\}\nonumber.
\end{eqnarray}

In Fig.\ref{F3}, we plot the Gaussian tripartite steering $\mathcal{G}^{\bar C \rightarrow\ (AC)}$, $\mathcal{G}^{(AC) \rightarrow\ \bar C}$, $\mathcal{G}^{B \rightarrow\ (A \bar C)}$, and $\mathcal{G}^{(A\bar C) \rightarrow\ B}$ as functions of the Hawking temperature parameter $\eta$ in the Schwarzschild black hole. From Fig.\ref{F3}, we can see that $\mathcal{G}^{(AC) \rightarrow\ \bar C}$ and $\mathcal{G}^{(A\bar C) \rightarrow\ B}$ first increase and then approach an asymptotic value with the growth of the $\eta$. However, the Hawking effect cannot generate $\mathcal{G}^{\bar C \rightarrow\ (AC)}$ and $\mathcal{G}^{B \rightarrow\ (A \bar C)}$,  indicating that the Hawking effect induces one-way steering between initially uncorrelated modes in the tripartite Gaussian systems.  In contrast, the Hawking effect can give rise to two-way steering in physically inaccessible bipartite Gaussian system \cite{ST4}.
These discoveries provide critical insights into the  redistribution principles governing Gaussian tripartite steering near the event horizon.

\section{ CONCLUSIONS  \label{GSCDGE}}
In this paper, we have analyzed the distribution of Gaussian tripartite steering in the context of a Schwarzschild black hole. Specifically, we explore the sharing of quantum steerability among four subsystems: modes $A$ and $B$, observed by Kruskal observers Alice and Bob, respectively; mode $C$, observed by a Schwarzschild observer Charlie, outside the event horizon;  mode $\bar{C}$, observed by anti-Charlie, inside the event horizon. Utilizing the Bogoliubov transformation between the Kruskal and Schwarzschild vacua, we derive the phase-space evolution of a Gaussian tripartite state under the influence of the Hawking effect.
Our results reveal that $\mathcal{G}^{(BC) \rightarrow\ A}$ initially decreases and subsequently approaches an asymptotic value as the Hawking temperature increases.
Notably, this indicates that physically accessible $2\rightarrow 1$ steerability  persists at all Hawking temperatures, in contrast to  $1\rightarrow 1$ steering
which undergoes ``sudden death" with  increasing Hawking  temperature \cite{ST4}.  This contrast highlights the superior robustness of Gaussian tripartite steering under the influence of the Hawking effect compared to bipartite steering. A particularly fascinating observation is the abrupt vanishing of quantum steering $\mathcal{G}^{A \rightarrow\ (BC)}$, which coincides with the onset of maximal steering asymmetry. This marks a pivotal transition from two-way to one-way steering for the tripartite Gaussian state in the black hole. These results offer a deeper understanding of how Gaussian tripartite steering adapts to extreme gravitational environment.
Ultimately, we conclude that the Hawking effect uniquely generates physically inaccessible $2 \rightarrow 1$ steering in tripartite Gaussian systems, manifesting as one-way steering between initially uncorrelated modes. However,
physically accessible Gaussian system is two-way steerable in curved spacetime. This work underscores the profound interplay between Gaussian tripartite steering and the Hawking effect, offering a richer understanding of quantum information dynamics in curved spacetime.

\begin{acknowledgments}
This work is supported by the National Natural
Science Foundation of China (Grant Nos. 12205133, 12475051, and 12035005);  the Special Fund for Basic Scientific Research of Provincial Universities in Liaoning under grant NO. LS2024Q002; and the science and technology innovation Program of Hunan Province under grant No.2024RC1050.
\end{acknowledgments}

\end{document}